\documentclass[letterpaper, 10 pt, conference]{ieeeconf}

\usepackage[top=1in, bottom=1in, left=0.75in, right=0.75in]{geometry}
\usepackage{caption}
\usepackage{cite}
\usepackage{amsmath,amssymb,amsfonts}
\usepackage{algorithmic}
\usepackage{graphicx}
\usepackage{textcomp}
\def\BibTeX{{\rm B\kern-.05em{\sc i\kern-.025em b}\kern-.08em
    T\kern-.1667em\lower.7ex\hbox{E}\kern-.125emX}}

\usepackage[utf8]{inputenc}
\usepackage{graphicx}               
\usepackage{amsmath,amssymb}
\usepackage[table,xcdraw]{xcolor}
\usepackage{listings}
\usepackage{xspace}
\usepackage{soul}
\usepackage{float}
\usepackage{booktabs}
\usepackage{wrapfig}

\usepackage{chngcntr}
\counterwithin{table}{subsection}
\usepackage{float}

\newcommand{\bx}{\boldsymbol{x}\xspace}
\newcommand{\bh}{\boldsymbol{h}\xspace}
\newcommand{\bu}{\boldsymbol{u}\xspace}
\newcommand{\bxdot}{\boldsymbol{\dot{x}}\xspace}

\newcommand{\bfz}{\boldsymbol{f_0}\xspace}
\newcommand{\bfpara}{\boldsymbol{f_{||}}\xspace}
\newcommand{\bfperp}{\boldsymbol{f_{\bot}}\xspace}
\newcommand{\bfphi}{\boldsymbol{f_{\phi}}\xspace}
\newcommand{\bO}{\boldsymbol{\mathcal{O}}\xspace}
\newcommand{\tbx}{\boldsymbol{\tilde{x}}\xspace}

\DeclareMathSymbol{\shortminus}{\mathbin}{AMSa}{"39}

\IEEEoverridecommandlockouts 
\title{\LARGE \bf
A Nonlinear Observability Analysis of Ambient Wind Estimation with Uncalibrated Sensors, Inspired by Insect Neural Encoding
}

\author{Floris van Breugel
\thanks{This work was partially supported by NIH (P20GM103650), AFRL (FA8651-20-1-0002), and the Sloan Foundation (FG-2020-13422)}
\thanks{F. van Breugel is a faculty member of Mechanical Engineering at the University of Nevada, Reno, USA
        {(email: \tt\footnotesize fvanbreugel@unr.edu}).}%
}

\begin{document}
\maketitle
\thispagestyle{empty}
\pagestyle{empty}


\begin{abstract}
Estimating the direction of ambient fluid flow is key for many flying or swimming animals and robots, but can only be accomplished through indirect measurements and active control. Recent work with tethered flying insects indicates that their sensory representation of orientation, apparent flow, direction of movement, and control is represented by a 2-dimensional angular encoding in the central brain. This representation simplifies sensory integration by projecting the direction (but not scale) of measurements with different units onto a universal polar coordinate frame. To align these angular measurements with one another and the motor system does, however, require a calibration of angular gain and offset for each sensor. This calibration could change with time due to changes in the environment or physical structure. The circumstances under which small robots and animals with angular sensors and changing calibrations could self-calibrate and estimate the direction of ambient fluid flow while moving remains an open question. Here, a methodical nonlinear observability analysis is presented to address this. The analysis shows that it is mathematically feasible to continuously estimate flow direction and perform regular self-calibrations by adopting frequent changes in course (or active prevention thereof) and orientation, and requires fusion and temporal differentiation of three sensory measurements: apparent flow, orientation (or its derivative), and direction of motion (or its derivative). These conclusions are consistent with the zigzagging trajectories exhibited by many plume tracking organisms, suggesting that perhaps flow estimation is a secondary driver of their trajectory structure.
\end{abstract}

\vspace{10pt}

\normalsize{}
\vspace{10pt}

\section{Introduction} \label{sec:introduction}
Determining the direction of ambient fluid flow is a crucial step for flying or swimming animals and robots during tasks such as tracking chemical plumes \cite{Baker2018}. Without access to stationary flow sensors, however, ambient flow direction cannot be directly measured \cite{schone2014}. Instead, agents moving in fluids can measure \textit{apparent flow}, corresponding to the vector sum of the ambient flow and motion induced flow (Fig. \ref{fig:fly}A). Decoupling apparent flow into its component parts is a straightforward vector subtraction if given absolute trajectory information in the same units as apparent flow. Many animals and the smallest robots, however, do not have access to absolute measurements from GPS, and cannot rely on Simultaneous Localization and Mapping (SLAM) due to the associated computational costs, which approach 10 Watts for monocular SLAM \cite{icra_2019_fastdepth}. 

Instead of relying on GPS and SLAM, some engineers \cite{Beyeler2009, Franceschini2014, Ohradzansky2018} have taken inspiration from insects, which rely on visual estimates of self-motion through optic flow \cite{Yang2018}, providing only relative information due to depth- \cite{van_Breugel_2014}, texture-\cite{OCarroll1996}, and state- \cite{Maimon2010} dependent response properties. Furthermore, variability due to manufacturing or development means some calibration is required \cite{Dhingra2020, Krause2019}. One calibration may be insufficient, however, since damage \cite{Foster2011} or temperature fluctuations \cite{Tatler2000, French1982} can alter responses of sensors and actuators. Thus, a mechanism for eliminating reliance on absolute measurement scales and continuous self-calibration would be helpful.

Flying insects have emerged as excellent model systems for studying how organisms track chemical plumes in moving fluids \cite{Carde2008, vanBreugel2014}, serving as inspiration for small robotic systems \cite{Anderson2020}. One strategy proposed to explain insects' ability to orient with respect to ambient wind, called visual anemotaxis \cite{Kennedy1940}, allows them to orient upwind by turning until the angle of perceived apparent wind is aligned with the angle of motion. Free and tethered flight behavior experiments with fruit flies, however, suggest that their turns are often open-loop and ballistic in nature, and not actively controlled throughout the turn itself  \cite{vanBreugel2012, Bender2006, Mongeau2017}. Thus, perhaps flies have more flexible algorithms for estimating ambient wind direction throughout their trajectory. Engineers have demonstrated solutions to this estimation challenge using externally calibrated sensor systems in Cartesian coordinates \cite{Zachariah2011,Rutkowski2011}. Insects, however, encode information in polar coordinates and must contend with shifting calibrations.

Recent discoveries using calcium imaging in tethered flying fruit flies, \textit{Drosophila melanogaster}, have found that a central part of their brain, the fan-shaped body and doughnut-shaped ellipsoid body, encodes 2-dimensional angular compass information such as head orientation from polarization \cite{Giraldo2018}, apparent wind angle from their antennae \cite{Okubo2020}, angular direction of motion from vision \cite{Weir2014, Green2017, Lyu}, and angular motor control commands \cite{Seelig2015}. Neuroanatomical studies suggest that similar sensory encodings exist in many invertebrates \cite{Thoen2017}. The physical overlap in the brain of the aforementioned sensory modalities presents an attractive hypothesis that perhaps they play a central role in insects' ability to estimate ambient wind direction. 

This manuscript presents a methodical nonlinear observability analysis of different sensor suites relevant to flying insects to understand the feasibility of estimating ambient wind direction and self-calibrating a sensorimotor system. Inspired by the 2-dimensional polar encoding found in the brains of insects, this analysis focuses on the 2-dimensional plane parallel to the ground, where the largest changes in wind direction are usually observed. Extensions to 3-dimensions are briefly discussed. To provide a concrete basis for the analysis and discussion, the term \textit{fly} is used throughout this paper. However, all of the concepts described here are equally applicable to any flying or swimming agent--biological or engineered--that has an interest in determining the direction of ambient fluid flow while moving.

\section{System Model} \label{sec:methods}

\begin{figure*}[ht] 
\centering
\includegraphics[width=0.8\textwidth]{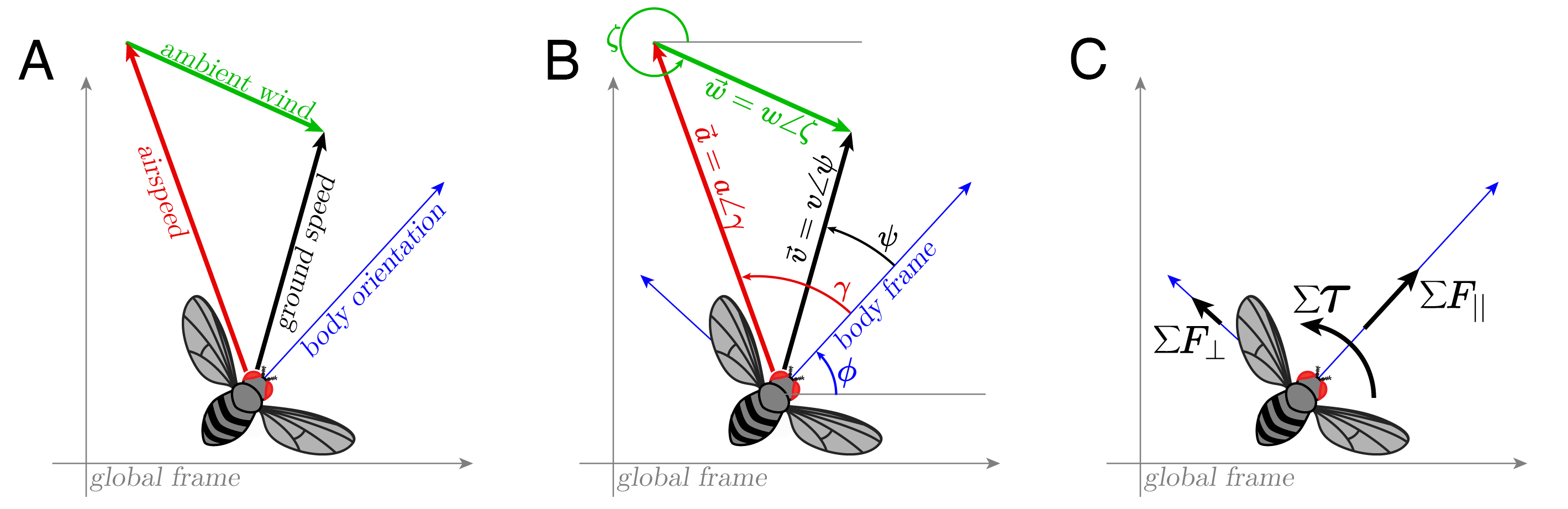}
\caption{2-dimensional geometric relationships of the states describing a flying fly in the presence of ambient wind.}
\label{fig:fly}
\end{figure*}

We begin with the equations of motion in the body frame of the fly (Fig. \ref{fig:fly}B), and model the fly's ability to control its flight as forces applied parallel $(||)$ and perpendicular $(\bot)$ to its body through its center of mass, as well as a torque to introduce angular accelerations (Fig. \ref{fig:fly}C). In addition to these controlled forces, there will also be linear and rotational aerodynamic drag forces. 
For this 2-dimensional description of the fly's dynamics, the full state describing the fly is $\bx = [v_{||}, v_{\bot}, \phi, \dot{\phi}, w, \zeta]^T$, where $v_{||}$ and $v_{\bot}$ describe the magnitude of the parallel and perpendicular velocity in the body coordinate frame, $\phi$ and $\dot{\phi}$ describe the angle of the fly relative to some global coordinate frame and its derivative (i.e. the fly's angular velocity), and $w$ and $\zeta$ describe the magnitude and direction of the wind in the global coordinate frame. Each of these states is a function of time,

\begingroup
\renewcommand*{\arraystretch}{1.2}

\begin{equation}
\label{eq:dynamics}
\bxdot = \\
\begin{bmatrix}
\dot{v}_{||}\\
\dot{v}_{\bot}\\
\dot{\phi}\\
\ddot{\phi}\\
\dot{w}\\
\dot{\zeta}
\end{bmatrix} =
\begin{bmatrix}
(F_{||} - D_{||})/m + v_{\bot}\dot{\phi}\\
(F_{\bot} - D_{\bot})/m - v_{||}\dot{\phi}\\
\dot{\phi}\\
\tau_{\phi} - D_{\phi} / I\\
\dot{w}\\
\dot{\zeta}
\end{bmatrix},
\end{equation}

\endgroup

where $F_{\bullet}$, $D_{\bullet}$, and $\tau_{\bullet}$ are the thrust forces, drag, and torque in the indicated directions, respectively. Here the subscript $\bullet$ indicates the collection of all relevant subscripts. The mass and moment of inertia are given by $m$ and $I$. Since $v_{||}$ and $v_{\bot}$ are defined in a moving reference frame, their derivatives must account for the derivative of the body coordinate frame, hence the $v_{\bullet}\dot{\phi}$ terms. For computational simplicity drag is treated as a linear function of airspeed\footnote{This simplification does not effect the local observability, since $C_{\bullet}$ may be a function of $\bx$ following the justification given for $k_{s\bullet}$ in Sec. \ref{sec:sensor_measurements}.}, 

\begin{equation}
\small
\label{eq:drag}
D_{\bullet}=
    \begin{cases}
      Drag_{||} = C_{||}a_{||}=C_{||}\cdot\big(v_{||}-w\cos(\phi-\zeta)\big), \\
      Drag_{\bot} = C_{\bot}a_{\bot}=C_{\bot}\cdot\big(v_{\bot}+w\sin(\phi-\zeta)\big), \\
      Drag_{\phi} = C_{\phi}\dot{\phi},
    \end{cases}
\end{equation}

where $C_{\bullet}$ represents a combined drag coefficient for the indicated direction. 

\subsection{Sensor measurements}
\label{sec:sensor_measurements}

The primary set of sensor measurements considered here are: a visual or polarization derived orientation with respect to world coordinates $(\phi)$ the angle of the airspeed (the negative of apparent flow) with respect to the body $(\gamma)$, and the angular course direction with respect to the body (i.e. drift) $(\psi)$. These angles can be defined relative to the vector components, 

\begingroup
\begin{equation}\label{eq:calibrated_h}
\begin{split}
\bh(\bx) = \begin{bmatrix}
\phi \\
\tan(\gamma) \\
\tan(\psi) \\
\end{bmatrix} &= \begin{bmatrix}
\phi \\
a_{\bot}/a_{||}  \\
v_{\bot}/v_{||}  \end{bmatrix},
\end{split}
\end{equation}
\endgroup

where the airspeed components are: 

\begingroup
\renewcommand*{\arraystretch}{2}
\begin{equation}
\begin{split}
a_{||} &= v_{||}-w\cos(\phi-\zeta), \\
a_{\bot} &= v_{\bot}+w\sin(\phi-\zeta). 
\end{split}
\end{equation}
\endgroup



Scenarios where instead of $\phi$ or $\psi$, only their derivatives are known, will also be considered. To model uncalibrated sensors we can introduce several unknown parameters $k_{s\bullet}$ that are assumed to remain constant in time for the duration over which the calibrations are performed,

\begingroup
\renewcommand*{\arraystretch}{1}
\begin{equation}
\label{eq:uncalibrated_sensors}
\begin{split}
\bh(\bx) = \begin{bmatrix}
& k_{s1}\cdot(\phi ) \\
& k_{s2}\cdot(a_{\bot}/a_{||}) + k_{s3} \\ 
& k_{s4}\cdot(v_{\bot}/v_{||}) + k_{s5}
\end{bmatrix}.
\end{split}
\end{equation}
\endgroup

The parameters $k_{s1}$,  $k_{s2}$, and $k_{s4}$ represent unknown sensor gains, and $k_{s3}$ and $k_{s5}$ represent unknown sensor offsets. Although this description implies that the sensors are linear with respect to these calibration constants, this is not necessarily the case. Since the observability calculations are done around some operating point $\bx_0$, if the system is observable, a calibration could be repeated at multiple operating points to determine the parameters as functions of the state. For a fly, a state-dependent calibration parameter could, for example, represent an asymmetric sensor, such as one that has a higher gain on the left side of the body compared to the right. With respect to the observability analysis, it is safe to ignore the fact that the parameters might actually be functions of the state (though solving for them will be more challenging).

To determine the nonlinear observability of the system, it is convenient to split the dynamics into the passive dynamics $(\bfz)$, and the active components $(\bfpara, \bfperp, \bfphi)$,

\begin{equation}\label{eq:xdotf}
\bxdot = \bfz + u_{||} \bfpara + u_{\bot} \bfperp + u_{\phi} \bfphi,
\end{equation}

where $u_{\bullet}$ are the control inputs. Relating this notation to the full equations of motion (Eqn. \ref{eq:dynamics}) yields:

\begingroup
\renewcommand*{\arraystretch}{1.2}
\begin{equation} \label{eq:2d_dynamics}
\scriptsize
\bxdot \! = \!\!\!\! \\
\stackrel{\mbox{$\bfz$}}{%
\begin{bmatrix}
\shortminus D_{||}/m + v_{\bot}\dot{\phi}\\
\shortminus D_{\bot}/m - v_{||}\dot{\phi}\\
\dot{\phi}\\
\shortminus D_{\phi}/I\\
\dot{w}\\
\dot{\zeta}
\end{bmatrix}%
}\  
\!\!\!\!+ u_{||}\!\!\!\\
\stackrel{\mbox{$\bfpara$}}{%
\begin{bmatrix}
k_{m1}/m \\
0 \\
0 \\
k_{m2}/I\\
0 \\
0 \\
\end{bmatrix}%
}\ 
\!\!\!\!+ u_{\bot}\!\!\!\\
\stackrel{\mbox{$\bfperp$}}{%
\begin{bmatrix} 
0 \\
k_{m3}/m \\
0 \\
0 \\
0 \\
0 \\
\end{bmatrix}%
}\
\!\!\!\!+u_{\phi}\!\!\!\\
\stackrel{\mbox{$\bfphi$}}{%
\begin{bmatrix} 
0 \\
0 \\
0 \\
k_{m4}/I\\
0 \\
0 \\
\end{bmatrix},%
}\
\end{equation}

\endgroup

where $k_{m\bullet}$ are parameters that describe how the control input $u_{\bullet}$ is transformed into an output force or torque (e.g. a model of the fly's muscle dynamics). Thus, $F_{||}=u_{||}k_{m1}$. The addition of $k_{m2}/I$ in $\bfpara$ models the possibility that one of the fly's wings might be damaged and a forward thrust command may induce an unexpected torque as well. As with the sensor calibration parameters, the $k_{m\bullet}$ parameters are assumed to be constant in time over the duration of the calibration, and although they may be functions of the state, it is unnecessary to consider this for the observability analysis. 

\section{Nonlinear Observability Analysis (a review)}

This section provides a brief applied overview of nonlinear observability, for a rigorous treatment see \cite{Hermann1977}. 
A nonlinear system is said to be globally observable when there is a one-to-one mapping from the time varying measurements  $\bh(\bx,\bu,t)$ and associated control inputs $\bu$ to the state $\bx$. 
More formally, if $\bh(\bx_1,\bu,t)=\bh(\bx_2,\bu,t)$ for all $\bu$ and $t$ implies that $\bx_1=\bx_2$, then the system is observable \cite{nijmeijer1990nonlinear}. 
For nonlinear systems this is difficult to prove in a \textit{global} sense. Instead, it is possible to prove \textit{local} observability in the neighborhood of a particular state $\tbx_0$ by comparing the number of unique equations given by the collection of the sensor measurements and their time derivatives ($\bO$) to the number of unknowns ($\tbx$: the state $\bx$ plus any parameters). An efficient approach for checking this condition is to determine if the Jacobian of $\bO$ with respect to $\tbx$ is full rank at $\tbx_0$. 

To begin, we calculate how the set of the sensor measurements, $\bh(\bx)$, changes with time, with and without active control. The derivative of $\bh(\bx)$ with respect to time is given by,

\begin{equation} \label{eq:dhdx}
\frac{\partial}{\partial t} \bh(\bx) = \frac{ \partial \bh}{\partial \bx} \cdot \frac{ \partial \bx}{\partial t} = \frac{ \partial \bh}{\partial \bx} \cdot \dot{\bx}.
\end{equation}

If $\bh$ consists of 3 sensor measurements, then for the 6-state system $\frac{ \partial \bh}{\partial \bx}$ will be a 3x6 matrix, and $\frac{ \partial \bh}{\partial \bx} \cdot \dot{\bx}$ will be a 3x1 matrix. Substituting Eqn. \ref{eq:xdotf} for $\bxdot$  in Eqn. \ref{eq:dhdx} yields:

\begin{equation}
\frac{\partial}{\partial t} \bh(\bx) = \frac{ \partial \bh}{\partial \bx} \cdot \big( \bfz + u_{||} \bfpara + u_{\bot} \bfperp + u_{\phi} \bfphi \big).
\end{equation}

If all of the control inputs are set to zero, then this is simply the directional derivative of the sensor measurements with respect to the passive dynamics. This is equivalent to calculating the \textit{Lie derivative} with respect to the passive dynamics, which provides compact notation:

\begin{equation}
\frac{\partial}{\partial t} \bh(\bx) \bigg\rvert_{u_{||}=u_{\bot}=u_{\phi}=0} = \frac{ \partial \bh}{\partial \bx} \cdot \bfz = L_{\bfz}\bh,
\end{equation}

where $L_{\bfz}\bh$ is the Lie derivative of $\bh$ in the $\bfz$ direction. If one of the control inputs is not zero, then it is possible that new information can be gathered by quantifying how $\bh$ changes during the controlled maneuver. For example, if $u_{||}$ is non-zero and other controls are held at zero, 

\begin{equation}
\begin{split}
\frac{\partial}{\partial t} \bh(\bx) \bigg\rvert_{u_{||}\neq0;\: u_{\bot}=u_{\phi}=0} &= \frac{ \partial \bh}{\partial \bx} \cdot \bigg(\bfz+u_{||}\bfpara\bigg) \\
&= \frac{ \partial \bh}{\partial \bx} \cdot \bfz + u_{||} \frac{ \partial \bh}{\partial \bx} \cdot \bfpara \\
&= L_{\bfz}\bh + u_{||} L_{\bfpara}\bh.
\end{split}
\end{equation}

Since the present goal is only to determine the observability, the calculations are simplified by setting $u_{||}=1$. If this process is repeated for all three control directions, then collecting all of the resulting terms from a first order time derivative of $\bh(\bx)$ yields the following collection, which is referred to as the \textit{observability Lie algebra},
\begin{equation}
\begin{split}
\bO^1 = \{& \bh,
L_{\bfz}\bh,
L_{\bfz}\bh+L_{\bfpara}\bh,\\
&L_{\bfz}\bh+L_{\bfperp}\bh,
L_{\bfz}\bh+L_{\bfphi}\bh \}.
\end{split}
\end{equation}
The superscript $1$ denotes the first order algebra, which includes only first order derivatives of the original sensor measurements. 
If $\bh$ contains 3 original sensor measurements, then the $\bO^1$ described above will be a list of 15 terms. 
To assess the observability, we calculate the Jacobian of $\bO$ with respect to all of the independent variables. Note that the set of independent variables is, at a minimum, equivalent to $\bx$, however, it may contain unknown parameters, as well as uncontrolled dynamics of the wind. This expanded set of independent variables will be referred to as $\tbx$. If the resulting Jacobian evaluated at some set of values $\tbx=\boldsymbol{\tilde{x}_0}$ is full rank, then the system is locally observable in the neighborhood of that particular $\boldsymbol{\tilde{x}_0}$. This calculation can be repeated for different values of $\boldsymbol{\tilde{x}_0}$ to determine states for which the system is not locally observable. 

In some cases, the full system may not be observable, whereas the key state(s) of interest could be. To assess the local observability of an individual state ($x_i$) we can compare the rank of the Jacobian of $\bO$ with the rank of the Jacobian of $\bO$ augmented with $x_i$. If the rank is unchanged after the augmentation, that indicates that $x_i$ was already contained within the original algebra $\bO$. Mathematically, if 
$\text{Rank} \big( \text{Jac}(\bO) \big \rvert_{\boldsymbol{\tilde{x}_0}} \big) =  \text{Rank} \big( \text{Jac}(\{\bO; x_i \}) \big \rvert_{\boldsymbol{\tilde{x}_0}} \big)$,
then $x_i$ is locally observable at the operating point of $\boldsymbol{\tilde{x}_0}$.

To simplify the observability calculations, $\bO^1$ can be written as $\{\bh,
     L_{\bfz}\bh,
     L_{\bfpara}\bh,
     L_{\bfperp}\bh,
     L_{\bfphi}\bh \}$.
This simplification corresponds to the fly comparing the derivative of its sensor measurements with and without the active control. 

In some cases, even further information can be gained by calculating the second order observability Lie algebra. In principle, $\bO^2$ contains all of the possible combinations of cross-terms between different control actuations, 
$\bO^2 = \{ \bh, 
   \bO^1, 
   L_{\bfz}\bO^1, 
   L_{\bfpara}\bO^1, 
   L_{\bfperp}\bO^1, 
   L_{\bfphi}\bO^1 \}$. 
For the scenarios considered here, it can be shown that including cross-terms other than $\bO^2 = \{\bh, \bO^1, L_{\bfz}\bO^1\}$ always introduces redundant complexity.


To efficiently explore different combinations of sensor measurements and control actuations, the aforementioned calculations were implemented using symbolic algebra in SymPy \cite{sympy}, and the Jacobians evaluated at values of $\boldsymbol{\tilde{x}_0}$ where each variable was assigned a unique (non-zero) prime number, unless otherwise noted.

\section{Analysis} \label{sec:results}

\subsection{Estimating wind direction using vision \& airspeed}

In the following three subsections we will walk through increasingly challenging scenarios to build intuition about which sensor measurements and active controls are necessary in which circumstances. At the end of each subsection a summary table is provided. Starting with the case of constant wind and no drag, a detailed example of the observability analysis is given. For the subsequent scenarios the details are omitted; but are provided in an open source supplement here: \texttt{https://github.com/florisvb/windobs}. 

\subsubsection{Calibrated sensorimotor systems in 2D}

\subsubsection*{\textit{Constant wind, no drag}}

In this first scenario, consider a completely calibrated sensorimotor system in constant wind, in the absence of air drag. Thus, the only forces and torques on the fly are those that it controls with its muscles. For this scenario, the only values that must be determined are the state itself, thus $\tilde{\bx} = [v_{||}, v_{\bot}, \phi, \dot{\phi}, w, \zeta]^T$. 


To simplify the mathematical expressions for readability, we can substitute the value of ``$1$" for each parameter, except $k_{m2}=0$ for simplicity, into Eqn. \ref{eq:2d_dynamics}.
With only the sensory measurement functions given by Eqn. \ref{eq:calibrated_h}
it is clear that the system is not fully observable, as there are only 3 unique terms in $\bh$, whereas $\tbx$ contains 6. Including the first order time derivative of $\bh$ such that $\bO^1=\{\bh, L_{\bfz} \bh\}$ adds three additional terms to the observability Lie algebra resulting in: 

\begingroup
\renewcommand*{\arraystretch}{2}
\scriptsize
\begin{equation} \label{eq:first_order_obs}
\begin{split}
&\bO^1 = \bigg\{\phi,\;\;
\frac{v_{\bot} + w \sin{\left (\phi - \zeta \right )}}{v_{||} - w \cos{\left (\phi - \zeta \right )}},\;\;
\frac{v_{\bot}}{v_{||}},\;\;
\dot{\phi},\;\;
- \dot{\phi} - \frac{v_{\bot}^{2} \dot{\phi}}{v_{||}^{2}},\\
& \frac{-\dot{\phi} \left(v_{||}^{2} - 2 v_{||} w \cos{\left (\phi - \zeta \right )} + v_{\bot}^{2} + 2 v_{\bot} w \sin{\left (\phi - \zeta \right )} + w^{2}\right)}{\left(v_{||} - w \cos{\left (\phi - \zeta \right )}\right)^{2}} \bigg\}.
\end{split}
\end{equation}
\endgroup


Since the observability of the system is determined at one particular value of $\tbx$ at a time, one of the new terms, $\dot{\phi}$, does provide new information. In other words, the inclusion of the $L_{\bfz} \bh$ term accounts for the fact that if $\phi$ is known at multiple time points, then $\dot{\phi}$ can also be determined. The last two terms, however, are simply combinations of the first 4. 
Only by including some element of active control in the form of thrust parallel or perpendicular to the body is enough new information gained such that $\zeta$ becomes observable. For example, if the fly applies thrust parallel to its body $\big(u_{||}\neq0\big)$ then the complete set of observations, $\bO^1 = \{\bh, L_{\bfz} \bh, L_{\bfpara} \bh \}$, adds the following terms to Eqn. \ref{eq:first_order_obs}:

\begingroup
\renewcommand*{\arraystretch}{2}
\begin{equation}
\label{eq:O_constantwind}
\bigg\{
0,\;\;
- \frac{v_{\bot} + w \sin{\left (\phi - \zeta \right )}}{\left(v_{||} - w \cos{\left (\phi - \zeta \right )}\right)^{2}},\;\;
- \frac{v_{\bot}}{v_{||}^{2}} \bigg\}.
\end{equation}
\endgroup

With the addition of these terms, the rank of the Jacobian of $\bO^1$ is now 6 (for most values of $\tbx$), making the system fully observable (most of the time). The unique terms for \{Eqn. \ref{eq:first_order_obs}; Eqn. \ref{eq:O_constantwind}\} are $\{1,2,3,4,8,9\}$. The system is also observable if $L_{\bfperp} \bh$ is included instead of $L_{\bfpara} \bh$, although the last three terms in $\bO^1$ will change. 

There are still certain conditions (states) for which $\zeta$ remains unobservable. For example, if the magnitude of the airspeed is zero (the fly is advected by the wind and therefore the airspeed angle is undefined) or the magnitude of the ground speed is zero (the fly is hovering and therefore the drift angle is undefined), then $\zeta$ remains unobservable. The latter case is resolved by incorporating inertial measurements (Sec. \ref{sec:inertial}).

Furthermore, the flies course direction (i.e. $\phi + \psi$, where $\psi=\tan^{-1}\big(v_{\bot}/v_{||}\big)$) must either change, or be controlled so as to actively counteract changes that would otherwise occur (due to changes in wind speed or direction, for example). In other words, to maintain observability, the fly cannot simply accelerate in a straight line, which would correspond to $u_{||}\neq0$, $v_{\bot}=0$, and $\dot{\phi}=0$. This requirement can be satisfied in a number of ways. (1) If $\dot{\phi}\neq0$, there will be a change in course due to the body rotation and the non-zero ground speed. (2) If $\dot{\phi}=0$, but $u_{||}\neq0$ and $v_{\bot}\neq0$, then $v_{\bot}/v_{||}$ will change. (3) Similarly, if $\dot{\phi}=0$, but $u_{\bot}\neq0$ and $v_{||}\neq0$, the system is observable if $v_{||}/v_{\bot}$ is included as a measurement instead of $v_{\bot}/v_{||}$. 

\begin{table}[H]
\footnotesize
\centering
\caption{Estimating wind direction $\zeta$ given 3 calibrated angular sensor measurements, and a calibrated motor system, requires applying thrust in either the forward or lateral direction, accompanied by a change in course (or orientation) and non-zero airspeed and ground speed. In dynamic wind, a $2^{nd}$ derivative must be computed. To estimate wind direction in the body-frame only ($\phi-\zeta$), the measurement $\phi$ may be replaced with $\dot{\phi}$.}
\label{tab:311}
\begin{tabular}{lcc}
\hline
\multicolumn{1}{c}{\textbf{}}                                              & \textbf{Description}                                                                                            & \textbf{Math Expr.}                                                                                                                    \\ \hline
\rowcolor[HTML]{EFEFEF} 
\textbf{Dynamics}                                                          & 2D, no drag                                                                                                     & Eqn. \ref{eq:2d_dynamics}, $D_{\bullet}=0$                                                                                             \\
\rowcolor[HTML]{FFFFFF} 
\textbf{\begin{tabular}[c]{@{}l@{}}Calibrated\\ parameters\end{tabular}}   & \begin{tabular}[c]{@{}c@{}}body,\\ motor,\\ \& sensor\end{tabular}                                              & \begin{tabular}[c]{@{}c@{}}$m$, $I$\\ $k_{m\bullet}$,\\ $k_{s\bullet}$\end{tabular}                                                    \\
\rowcolor[HTML]{EFEFEF} 
\textbf{\begin{tabular}[c]{@{}l@{}}Uncalibrated\\ parameters\end{tabular}} & None                                                                                                            & None                                                                                                                                   \\
\rowcolor[HTML]{FFFFFF} 
\textbf{Sensors $\bh(\bx)$}                                                & \begin{tabular}[c]{@{}c@{}}orientation,\\ visual angle,\\ airspeed angle\end{tabular}                               & \begin{tabular}[c]{@{}c@{}}$\phi$, \\ $v_{\bot}/v_{||}$,\\ $a_{\bot}/a_{||}$\end{tabular}                                              \\
\rowcolor[HTML]{EFEFEF} 
\textbf{\begin{tabular}[c]{@{}l@{}}Required\\ actuations\end{tabular}}     & \begin{tabular}[c]{@{}c@{}}forward thrust\\ - - OR - -\\ lateral thrust\end{tabular}                            & \begin{tabular}[c]{@{}c@{}}$u_{||}\neq0$\\ - - OR - -\\ $u_{\bot}\neq0$\end{tabular}                                                   \\
\textbf{\begin{tabular}[c]{@{}l@{}}Required\\ state\end{tabular}}          & \begin{tabular}[c]{@{}c@{}}non-zero ground speed, \\ \& non-zero airspeed, \\ \& change in course/orientation\end{tabular} & \begin{tabular}[c]{@{}c@{}}$v^2_{||}+v^2_{\bot}\neq0$,\\ \& $a^2_{||}+a^2_{\bot}\neq0$,\\ \& ($\dot{\psi}\neq0$ or $\dot{\phi}\neq0$)\end{tabular} \\ \hline
\end{tabular}
\end{table}

\subsubsection*{\textit{Dynamic wind, no drag}}

Now we will relax the earlier assumption of constant wind to see how the requirements for observability change when the wind is allowed to vary as a function of time. It is now necessary to also estimate $\dot{w}$ and $\dot{\zeta}$, thus $ \tilde{\bx} = [v_{||}, v_{\bot}, \phi, \dot{\phi}, w, \zeta, \dot{w}, \dot{\zeta}]^T$. As a first step, we check whether the conditions for observing $\zeta$ with constant wind are sufficient, i.e. is the system observable given $\bO^1 = \{\bh, L_{\bfz} \bh, L_{\bfpara} \bh \}$. In this case, $\text{Rank}(\text{Jac}(\bO^1))=7$, whereas $\tbx$ contains 8 terms. By augmenting $\bO^1$ with individual states and comparing the Rank of the Jacobian of these augmented algebras with the original it can be shown that $v_{||}, v_{\bot}, \zeta$ and $w$ are observable, whereas $\dot{w}$ and $\dot{\zeta}$ are not. However, by including second order terms, the system can be shown to be fully (locally) observable. 


\subsubsection*{\textit{Dynamic wind, with drag}}

A real fly experiences aerodynamic drag forces on its body that will oppose its airspeed, introducing passive deceleration. 
With the addition of drag, estimating $\zeta$ does not require any thrust, contrary to the previous scenario. Instead, only a single derivative of the measurements is required, or a $2^{nd}$ derivative if the wind is dynamic and $\dot{\zeta}$ estimates are desired. This is because the drag forces will introduce the necessary changes in $v_{||}$ and $v_{\bot}$ for the system to become observable. In practice, it is likely more efficient to apply active controls rather than rely on the drag. 

\subsubsection*{\textit{Eliminating the global reference}}

The global angular reference $\phi$ is necessary for $\zeta$ to be observable in global coordinates. Although many insects have access to some measure of $\phi$ through polarization sensors, they also have access to angular velocity measurements ($\dot{\phi}$) through either the halteres \cite{Pringle1948,Dickerson2019}, antennae \cite{Sane2007}, or wings \cite{Mohren2019} without needing to differentiate their measurement of $\phi$. By including $\dot{\phi}$ in $\bh$ instead of $\phi$, the wind direction in body-coordinates ($\phi-\zeta$) remains observable.

\subsubsection{Uncalibrated sensorimotor systems in 2D}

Thus far we have assumed that the fly somehow has knowledge of the parameters related to its body ($m, I, C_{\bullet}$) and motor systems ($k_{m\bullet}$). It is, however, likely that flies must calibrate their body and muscle parameters at least once, if not continuously. Likewise, we have thus far assumed that the parameters of the sensory system ($k_{s\bullet}$) are perfectly calibrated. For example, we have assumed that when the fly's body is aligned with its airspeed ($\gamma=0; a_{\bot}/a_{||}=0$), then its \textit{measurement} of $\gamma$ is indeed equal to zero. This might not be the case due to damage or developmental asymmetries. The same arguments can be made for the visually determined measure of $\psi$. In this section we will build on the previous analysis to determine (1) which sensory measurements and active maneuvers are necessary to estimate $\zeta$ given a completely uncalibrated sensorimotor system, and (2) what is necessary to calibrate all of the unknown parameters. 

\subsubsection*{\textit{Estimating $\zeta$ with an uncalibrated sensorimotor system}}

Analyzing the dynamics described by Eqn. \ref{eq:2d_dynamics} and the uncalibrated measurements given by Eqn. \ref{eq:uncalibrated_sensors}, it is unsurprising  that the previously successful approach--$\bO^2 = \{\bh, L_{\bfz} \bh, L_{\bfpara} \bh, L_{\bfz} L_{\bfpara} \bh \}$--is now insufficient ($\text{Rank}(\text{Jac}(\bO^2))=13$; $\text{Rank}(\text{Jac}(\bO^2; \zeta))=14$). However, by applying thrust in both the perpendicular \textit{and} parallel directions, $\zeta$ does become observable with  

\begin{equation}
\label{eq:uncalibrated_Otwo}
\begin{split}
\bO^2 = \{&\bh, L_{\bfz} \bh, L_{\bfpara} \bh, L_{\bfperp} \bh, \\
& L_{\bfz} L_{\bfpara} \bh, L_{\bfz} L_{\bfperp} \bh \},
\end{split}
\end{equation}
yielding $\text{Rank}(\text{Jac}(\bO^2))=17$ and $\text{Rank}(\text{Jac}(\bO^2; \zeta))=17$. Contrary to the prior cases, in constant wind a second derivative is still necessary to determine $\zeta$. There are, however, a total of 19 terms in $\tbx=[v_{||}, v_{\bot}, \phi, \dot{\phi}, w, \zeta, \dot{w}, \dot{\zeta}, C_{\bullet}, k_{s\bullet}, m, k_{m1}, k_{m3}]^T,$
indicating that while $\zeta$ can be observed, several other terms cannot. The unobservable terms include $\{w, v_{\bullet}, C_{\phi}, k_{m1}, k_{m3}\}$. Also notable is that the body parameters that determine rotational control ($I, k_{m2,4}$) are not found in $\bO^2$, and are therefore omitted from $\tbx$ and unobservable. Both points are addressed in Sec. \ref{sec:entire_sensorimotor}. 

As in the calibrated case, $\zeta$ is not observable for all values of $\tbx$: both airspeed and ground speed must be non-zero, otherwise the angles are undefined. Of particular interest is that $\dot{\phi}$ must be non-zero, and the alternative methods for changing course that worked in the calibrated case are insufficient. Finally, for calibration constants to be observable in \textit{either} constant or dynamic wind scenarios, the second order terms of $\bO^{2}$ are needed.

\subsubsection*{\textit{Calibrating the entire sensorimotor system}}
\label{sec:entire_sensorimotor}

To calibrate all of the parameters that appear in Eqn. \ref{eq:2d_dynamics} and Eqn. \ref{eq:uncalibrated_sensors} requires that they all appear in the observability Lie algebra. The simplest solution is to include $\dot{\phi}$ in $\bh$ and apply torque. 
By adding a new uncalibrated sensor, $k_{s6}\dot{\phi}+k_{s7}$, to $\bh$ given in Eqn. \ref{eq:uncalibrated_sensors}, the complete list of terms now includes $\tbx=[v_{||}, v_{\bot}, \phi, \dot{\phi}, w, \zeta, \dot{w}, \dot{\zeta},
      C_{\bullet}, k_{s1-7}, k_{m1-4}, m, I]^T$.
Furthermore, each of these terms appears in the second order algebra given by $\bO^2 = \{\bh, L_{\bfz} \bh, 
L_{\bfpara} \bh, L_{\bfz} L_{\bfpara} \bh, 
L_{\bfperp} \bh, \\
L_{\bfz} L_{\bfperp} \bh, 
L_{\bfphi} \bh, L_{\bfz} L_{\bfphi} \bh \}$. Even after making this change, however, some of the body parameters still cannot be determined: $\{k_{m1}, k_{m3}, m, I\}$. There are two reasons for this. First, $m$ and $I$ always appear together with another body or motor parameter. The solution is simple: $m$ and $I$ must be absorbed into the definitions of the parameters that they appear with. 
After doing so, the only body or motor parameters that cannot be determined are $k_{m1}$ and $k_{m3}$. 

Parameters $k_{m1}$ and $k_{m3}$ and states $w$ and $v_{\bullet}$ cannot be determined with an uncalibrated sensorimotor because they all relate to the scaling of the observability problem. If $k_{m1}$ were known, then known motor commands $u_{||}$ would translate into known accelerations with meaningful units. To determine these body parameters and states, at least one properly scaled sensory measurement (in meaningful units) is needed. 

\begin{table}[H]
\footnotesize
\centering
\caption{Estimating $\zeta$ given 3 uncalibrated angular sensor measurements and body/motor parameters, requires applying thrust in both the forward and lateral directions, accompanied by a change in orientation and non-zero airspeed and ground speed. To determine wind direction in body-coordinates ($\phi-\zeta$), the measurement $k_{s1}\phi$ can be replaced with $k_{s1}\dot{\phi}+k_{s0}$. Contrary to the calibrated case, constant wind also requires $2^{nd}$ order terms.}
\label{tab:312}
\begin{tabular}{lcc}
\hline
\multicolumn{1}{c}{\textbf{}}                                              & \textbf{Description}                                                                                                & \textbf{Math Expr.}                                                                                                         \\ \hline
\rowcolor[HTML]{EFEFEF} 
\textbf{Dynamics}                                                          & 2D, including drag                                                                                                  & Eqn. \ref{eq:2d_dynamics}                                                                                                   \\
\rowcolor[HTML]{FFFFFF} 
\textbf{\begin{tabular}[c]{@{}l@{}}Calibrated\\ parameters\end{tabular}}   & None                                                                                                                & None                                                                                                                        \\
\rowcolor[HTML]{EFEFEF} 
\textbf{\begin{tabular}[c]{@{}l@{}}Uncalibrated\\ parameters\end{tabular}} & \begin{tabular}[c]{@{}c@{}}body,\\ motor,\\ \& sensor\end{tabular}                                                  & \begin{tabular}[c]{@{}c@{}}$m$, $I$, $C_{\bullet}$,\\ $k_{m\bullet}$,\\ $k_{s\bullet}$\end{tabular}                         \\
\rowcolor[HTML]{FFFFFF} 
\textbf{Sensors $\bh(\bx)$}                                                & \begin{tabular}[c]{@{}c@{}}orientation,\\ airspeed angle,\\ visual angle\end{tabular}                               & \begin{tabular}[c]{@{}c@{}}$k_{s1}\phi$, \\ $k_{s2}a_{\bot}/a_{||}+k_{s3}$,\\ $k_{s4}v_{\bot}/v_{||}+k_{s5}$\end{tabular}   \\
\rowcolor[HTML]{EFEFEF} 
\textbf{\begin{tabular}[c]{@{}l@{}}Required\\ actuations\end{tabular}}     & \begin{tabular}[c]{@{}c@{}}forward thrust\\ - - AND - -\\ lateral thrust\end{tabular}                               & \begin{tabular}[c]{@{}c@{}}$u_{||}\neq0$\\ - - AND - -\\ $u_{\bot}\neq0$\end{tabular}                                       \\
\textbf{\begin{tabular}[c]{@{}l@{}}Required\\ state\end{tabular}}          & \begin{tabular}[c]{@{}c@{}}non-zero ground speed, \\ \& non-zero airspeed, \\ \& change in orientation\end{tabular} & \begin{tabular}[c]{@{}c@{}}$v^2_{||}+v^2_{\bot}\neq0$,\\ \& $a^2_{||}+a^2_{\bot}\neq0$,\\ \& $\dot{\phi}\neq0$\end{tabular} \\ \hline
\end{tabular}
\end{table}

\subsection{Estimating wind direction using acceleration \& airspeed}
\label{sec:inertial}

Although most organisms use vision to help them orient relative to fluid flow, some animals perform these behaviors in environments with poor visual contrast, such as when flying at night, over low contrast terrain, or swimming in murky waters. In this section we will find that an inertial sense of linear acceleration in polar coordinates (measured using the same sensors as angular velocities \cite{Lopez2020}) can serve in lieu of a visually determined angle of motion. For these calculations we simply substitute $k_{s4}\frac{\dot{v}_{\bot}}{\dot{v}_{||}}+k_{s5}$ for $k_{s4}\frac{v_{\bot}}{v_{||}}+k_{s5}$ from Eqn. \ref{eq:uncalibrated_sensors} and recalculate $\bO^2$. It is, however, necessary to substitute in the expressions for $\dot{v}_{\bullet}$ such that these new sensor measurements are described entirely by the state of the system,

\begingroup
\renewcommand*{\arraystretch}{2}
\begin{equation}
\small
\label{eq:vdot_eqns}
\begin{split}
\dot{v}_{||} &= v_{\bot}\dot{\phi} - C_{||}/m\big(v_{||}-w\cos(\phi-\zeta)\big) + u_{||}k_{m1}/m,\\
\dot{v}_{\bot} &= -v_{||}\dot{\phi} - C_{\bot}/m\big(v_{\bot}+w\sin(\phi-\zeta)\big) + u_{\bot}k_{m3}/m.
\end{split}
\end{equation}
\endgroup

Two of the control inputs, $u_{||}$ and $u_{\bot}$, appear in these expressions, and since they are known by default (they can be thought of as efferent copies), they can (and must) be included as measurements, resulting in: 
\begingroup
\renewcommand*{\arraystretch}{1.5}
\begin{equation}
\label{eq:uncalibrated_sensors_inertial}
\begin{split}
\bh(\bx) = 
\big[&k_{s1} \phi,\;
k_{s2}\cdot\big(a_{\bot}/a_{||}\big) + k_{s3},\\
&k_{s4}\cdot\big(\dot{v}_{\bot}/\dot{v}_{||}\big) + k_{s5},\;
u_{||},\;
u_{\bot}\big]^{T}.
\end{split}
\end{equation}
\endgroup

The previous approach, which worked with an uncalibrated visual angle, does not work here $\big( \text{Rank}(\text{Jac}(\bO^2))=19$ and $\text{Rank}(\text{Jac}(\bO^2; \zeta))=20$, where $\bO^2$ is given by Eqn. \ref{eq:uncalibrated_Otwo}$ \big)$. The key term that prevents the system from being observable is $k_{s5}$. If $k_{s5}$ is known, then $\zeta$ is observable; without $k_{s5}$, wind direction in the body-frame ($\phi-\zeta$) is still observable if $k_{s1}\phi$ is replaced by $k_{s1}\dot{\phi}+k_{s0}$ (or $L_{\bfpara}L_{\bfz}\bh$ and $L_{\bfperp}L_{\bfz}\bh$ are included in $\bO^2$). 
If the calibration terms are temporally stable over significant periods of time it would also be possible to use the visual system to calibrate $k_{s5}$ when both modalities are available. 

Once again, there are certain states for which $\zeta$ cannot be determined. The airspeed must still be non-zero and the fly must still rotate its body, however, the ground speed may now be zero, although either $\dot{v}_{||}$ or $\dot{v}_{\bot}$ must be non-zero. 
This approach would allow a hovering fly to estimate $\zeta$ by rotating and continuously adjusting $\dot{v}_{||,\bot}$ to remain in place.

\begin{figure*}[ht] 
\centering
\includegraphics[width=1\textwidth]{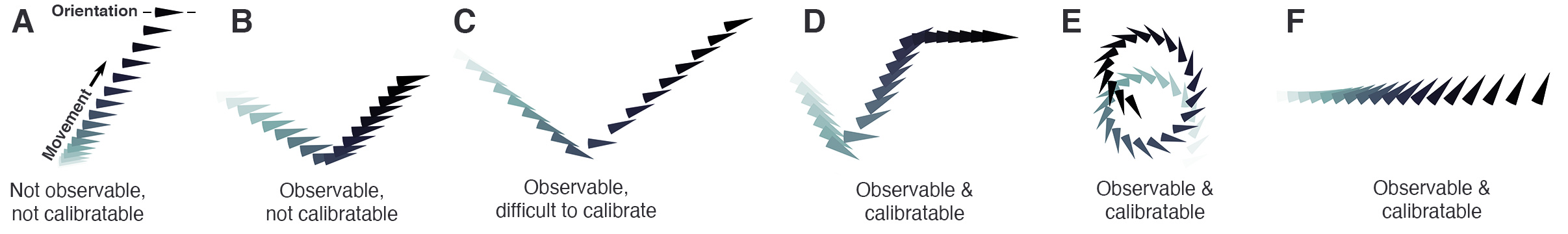}
\caption{Example trajectories in constant wind categorized according to the potential for observing wind direction ($\zeta$) for the wind-vision case with or without sensorimotor calibrations (Eqns. \ref{eq:calibrated_h} \& \ref{eq:uncalibrated_sensors}, respectively). Calibratable: the uncalibrated system is observable. Color (light to dark) encodes increasing time, $\zeta=\pi/2$, and arrows indicate body orientation. (A) With no change in direction or orientation, $\zeta$ is unobservable, even if $u_{||}\neq0$ \& $u_{\bot}\neq0$. (B) Changes in direction makes $\zeta$ observable, but without changes in orientation, sensorimotor parameters cannot be calibrated. (C) Incorporating changes in orientation with a turn, in principle, allows for calibration. However, the requirement for $2^{nd}$ order derivatives makes calibration with a single rapid turn challenging. (D-E) Two or more turns, or continuous changes in orientation and direction, are ideal. (F) By changing orientation and controlling $u_{||, \bot}$ for a constant course, calibration is also possible. In dynamic wind, one solution is for trajectory changes to occur at a sufficiently fast rate (see Fig. \ref{fig:ukf}).}
\label{fig:summary}
\end{figure*}

\begin{table}[H]
\footnotesize
\centering
\caption{Estimating the wind direction $\zeta$ with the direction of acceleration instead of velocity, and uncalibrated sensors, requires applying thrust in both the forward and lateral directions, accompanied by a change in orientation, non-zero airspeed, and non-zero acceleration. One sensory parameter, $k_{s5}$, cannot be self-calibrated in this case unless $k_{s6}v_{\bot}/v_{||}+k_{s7}$ is also included, or $\zeta$ is only estimated in body-coordinates by replacing $k_{s1}\phi$ with $k_{s1}\dot{\phi}+k_{s0}$ (or including additional terms in $\bO^2$).}
\label{tab:321}
\begin{tabular}{lcc}
\hline
\multicolumn{1}{c}{\textbf{}}                                              & \textbf{Description}                                                                    & \textbf{Math Expr.}                                                                                                                   \\ \hline
\rowcolor[HTML]{EFEFEF} 
\textbf{Dynamics}                                                          & 2D, including drag                                                                      & Eqn. \ref{eq:2d_dynamics}                                                                                                             \\
\rowcolor[HTML]{FFFFFF} 
\textbf{\begin{tabular}[c]{@{}l@{}}Calibrated\\ parameters\end{tabular}}   & None                                                                                    & $k_{s5}$                                                                                                                              \\
\rowcolor[HTML]{EFEFEF} 
\textbf{\begin{tabular}[c]{@{}l@{}}Uncalibrated\\ parameters\end{tabular}} & \begin{tabular}[c]{@{}c@{}}body,\\ motor,\\ \& sensor\end{tabular}                      & \begin{tabular}[c]{@{}c@{}}$m$, $I$, $C_{\bullet}$,\\ $k_{m\bullet}$,\\ $k_{s1-4}$\end{tabular}                                       \\
\rowcolor[HTML]{FFFFFF} 
\textbf{Sensors $\bh(\bx)$}                                                & \begin{tabular}[c]{@{}c@{}}orientation,\\ airspeed angle,\\ inertial angle\end{tabular} & \begin{tabular}[c]{@{}c@{}}$k_{s1}\phi$, \\ $k_{s2}a_{\bot}/a_{||}+k_{s3}$,\\ $k_{s4}\dot{v}_{\bot}/\dot{v}_{||}+k_{s5}$\end{tabular} \\
\rowcolor[HTML]{EFEFEF} 
\textbf{\begin{tabular}[c]{@{}l@{}}Required\\ actuations\end{tabular}}     & \begin{tabular}[c]{@{}c@{}}forward thrust\\ - - AND - -\\ lateral thrust\end{tabular}   & \begin{tabular}[c]{@{}c@{}}$u_{||}\neq0$\\ - - AND - -\\ $u_{\bot}\neq0$\end{tabular}                                                 \\
\textbf{\begin{tabular}[c]{@{}l@{}}Required\\ state\end{tabular}}          & \begin{tabular}[c]{@{}c@{}}non-zero acceleration, \\ \& non-zero airspeed, \\ \& change in orientation\end{tabular}  & \begin{tabular}[c]{@{}c@{}}$\dot{v}^2_{||}+\dot{v}^2_{\bot}\neq0$,\\ \& $a^2_{||}+a^2_{\bot}\neq0$,\\ \& $\dot{\phi}\neq0$\end{tabular}                                            \\ \hline
\end{tabular}
\end{table}

\newpage
\section{Discussion}
\begin{figure}[ht] 
\centering
\includegraphics[width=0.48\textwidth]{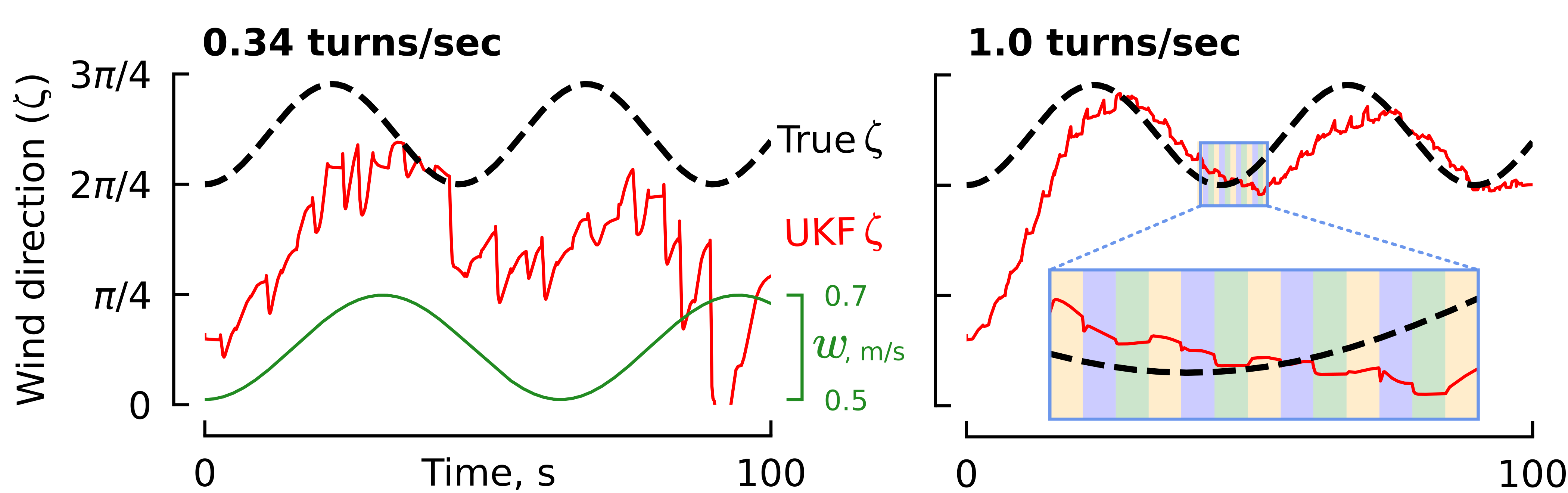}
\caption{Applying a square-root Unscented Kalman Filter (UKF) to the calibrated wind-vision case (Eqn. \ref{eq:calibrated_h}) for a trajectory similar to Fig. \ref{fig:summary}D, but with either 34 or 100 turns, nearly converges on the correct value of $\zeta$ given sufficiently frequent turns, even when both $\zeta$ and $w$ (wind magnitude) are changing in time according to unknown dynamics. Inset: magnified trace showing timing of left/right/straight turns. No added sensor noise, simulation time-step $=0.01$, UKF covariances: $R=10^{-7}I_3$, $Q=10^{-10}I_6$.}
\label{fig:ukf}
\end{figure}

The analysis presented here demonstrates the mathematical potential for flying or swimming agents to determine the direction of ambient flow using uncalibrated angular measurements of apparent flow, orientation, and direction of motion (or its derivative). All three measurements are \textit{required}. To estimate flow direction in the body-frame, angular velocity may be substituted for orientation. Solutions only exist, however, for trajectories that, broadly speaking, incorporate some active change in direction and orientation along with non-zero airspeed and ground speed (Fig. \ref{fig:summary}). The need for terms like $L_{\bfpara}\bh$ in $\bO$ indicates the requirement for comparing sensor measurements across time (a derivative) during these active maneuvers. Actually constructing a reliable nonlinear observer to estimate $\zeta$ is, however, not trivial. For example, applying a square-root Unscented Kalman Filter \cite{VanderMerwe} to the measurements and dynamics given by Eqns. \ref{eq:calibrated_h}, \ref{eq:2d_dynamics} for actively controlled trajectories like Fig. \ref{fig:summary}D requires \textit{many} turns before converging, even without sensor noise (Fig. \ref{fig:ukf}), even though wind direction should be observable after a single turn. The same filter completely fails on more challenging scenarios. For example, in the calibrated wind-inertial case there are frequent divide-by-zero issues when $\dot{v}_{||}$ changes sign, whereas in the uncalibrated wind-vision case the UKF has trouble likely because this scenario is not \textit{fully} observable. A closer look at insect sensory processing (e.g. \cite{Currier2020}) and strategies for control may offer insight for building a more reliable bespoke observer.

Real-world flow estimation is not restricted to 2-dimensions. However, with the approach presented here, it can be shown that the 3D case is also observable, provided that two additional sensory inputs are available: the angle for out-of-plane fluid flow, and an angle describing the ground plane orientation. Insects likely solve the latter by maintaining a stable head orientation with respect to the world using visual reflexes for gaze stabilization \cite{Huston2008}, and altitude control \cite{Straw2010}. The bilateral symmetry in the neural circuitry flies use for estimating the planar direction of flow \cite{Suver2019}, however, does not suggest mechanisms for measuring out-of-plane flow. Near the ground, vertical wind magnitude is attenuated compared to planar wind, so perhaps estimating vertical flow is less critical. Insects do, however, exhibit vertical motion during plume tracking \cite{vanBreugel2014,Willis2013}, so further investigation is warranted. 

Although ambient wind estimation does not require estimating the absolute scale factor, some behaviors such as estimating distance to a landing target \cite{van_Breugel_2014} or distance flown, would greatly benefit from an estimation of scale. Perhaps flying insects rely on a visual scale factor determined while walking \cite{Krause2019}, or controlled movements of the antennae to calibrate the scale of their flow sensors \cite{Lopez2020, RoyKhurana2016}. 

The requirement of frequent changes in direction for observability and calibratability is consistent with the crosswind zigzagging trajectories adopted by most chemical plume tracking organisms \cite{Baker2018}, suggesting a hypothesis that casting may aid in flow direction estimation in addition to facilitating plume encounters. Insect sized robots tasked with plume tracking might also benefit from constrained trajectories similar to those exhibited by insects. For larger machines carrying calibrated sensors, this analysis lays the groundwork for developing visual-wind and wind-inertial state estimation, which if combined with visual-inertial methods (i.e. \cite{icra_2019_fastdepth}) would increase robustness through redundancy. 

\section*{Acknowledgements}
I am grateful for feedback on the manuscript from N. Brace, A. Rutkowski, R. Tung, R. Murray, and K. Nagel.

\bibliography{references} 
\bibliographystyle{ieeetr}


\end{document}